\documentclass[]{INCLUDES/llncs}
\usepackage{amsmath}
\usepackage{graphicx}
\usepackage{subfig}
\usepackage{url}
\usepackage{verbatim} 
\usepackage{listings}
\lstnewenvironment{code}
    {\lstset{}%
      \csname lst@SetFirstLabel\endcsname}
    {\csname lst@SaveFirstLabel\endcsname}
\newtheorem{prop}{Proposition}

\def \bscale1 {1.00}
\def \bscale {0.20}
\def \hscale {0.20}

\newcommand{\FIG}[3]{
\begin{figure}[htbp]
\centering
\scalebox{\bscale1}{\includegraphics*[bb=0pt 0pt 800pt 800pt]{./figs/#3.png}}
\caption{#2}
\label{#1}
\end{figure}
}

\newcommand{\HFIGS}[6]{
\begin{figure}[htbp]
  \centering
  \subfloat[#3]{
    {\scalebox{\hscale}{\includegraphics*[bb=0pt 0pt 800pt 800pt]
    {./figs/#5.png}}}} 
  \subfloat[#4]{
    {\scalebox{\hscale}{\includegraphics*[bb=0pt 0pt 800pt 800pt]
    {./figs/#6.png}}}}
  \caption{#2}
  \label{#1}
\end{figure}
}

\begin{document}

\title{
   Executable Set Theory and Arithmetic Encodings in Prolog
}

\author{Paul Tarau}
\institute{
   Department of Computer Science and Engineering\\
   University of North Texas\\
   %P.O. Box 311366\\
   %Denton, Texas 76203\\
   {\em E-mail: tarau@cs.unt.edu}
}

\maketitle

\date{}

\begin{abstract}
The paper is organized as a self-contained literate Prolog program that
implements elements of an executable finite set theory with focus on
combinatorial generation and arithmetic encodings. 
The complete SWI-Prolog tested code is available at
\url{http://logic.csci.unt.edu/tarau/research/2008/pHFS.zip}.

First, ranking and unranking functions for some ``mathematically elegant" data
types in the universe of Hereditarily Finite Sets with Urelements are provided, 
resulting in arithmetic encodings for powersets, hypergraphs, ordinals and
choice functions. 

After implementing a digraph representation of 
Hereditarily Finite Sets we define {\em decoration functions} 
that can recover well-founded sets 
from encodings of their associated acyclic digraphs.

We conclude with an encoding of arbitrary digraphs and discuss 
a concept of duality induced by the set membership relation.

In the process, we uncover the surprising possibility of internally sharing 
isomorphic objects, independently of their language level types and meanings.

\end{abstract}

\section{Introduction}

This paper is an exploration with logic programming tools of interesting
executable aspects of finite set theory, with focus on natural number encodings
of various set-related data objects. Such encodings can be traced back to the
G\"{o}del numberings used to encode formulae as natural
numbers in G\"{o}del's famous incompleteness theorems. 
Their bijective variants are known as {\em ranking/unranking functions} in the
field of combinatorial generation, where they are used to iterate over 
various combinatorial objects. Other typical uses are in
the uniform generation of random objects of a given type and in designing
succinct representations for data compression purposes.

The practical expressiveness of logic programming languages (in particular
Prolog) are put at test in the process. This paper is part
of a larger effort to cover in a declarative programming 
paradigm, arguably more elegantly, some fundamental combinatorial generation 
algorithms along the lines of \cite{knuth06draft}.

The paper is organized as follows: section \ref{ack} introduces
Ackermann's encoding in the more general case when {\em urelements} 
are present and shows an encoding for hypergraphs as a particular case. 
Section \ref{functors} gives examples of
transporting common set and natural number operations from one side to the other.
After discussing some classic pairing functions, section \ref{pairings} 
introduces a new pairing/unpairing operation on natural numbers and applies them
in section \ref{ord} to obtain encodings of powersets, ordinals and choice
functions. Section \ref{graphs} discusses graph representations 
and {\em decoration functions} on Hereditarily Finite Sets (\ref{graphrep}), 
and provides encodings for directed acyclic graphs (\ref{digraphs}). 
Sections \ref{related} and \ref{concl} discuss related work, 
future work and conclusions.

\section{Hereditarily Finite Sets and the Ackermann Encoding}

While the Universe of Hereditarily Finite Sets is best known as a model 
of the Zermelo-Fraenkel Set theory with the Axiom of Infinity 
replaced by its negation \cite{finitemath}, 
it has been the object of renewed practical interest in various fields, 
from representing structured data in databases \cite{DBLP:conf/foiks/LeontjevS00} 
to reasoning with sets and set constraints in a Logic Programming framework 
\cite{dovier00comparing,DBLP:journals/tplp/PiazzaP04,DBLP:conf/cav/DovierPP01}.

\subsection{Ackermann's Encoding} \label{ack}

The Universe of Hereditarily Finite Sets is built from the empty set (or a set of {\em Urelements}) 
by successively applying powerset and set union operations.

A surprising bijection, discovered by Wilhelm Ackermann in 1937 \cite{ackencoding,kaye07} 
maps Hereditarily Finite Sets ($HFS$) to Natural Numbers ($Nat$):

\vskip 0.5cm
$f(x)$ = {\tt if} $x=\{\}$ {\tt then} $0$ {\tt else} $\sum_{a \in x}2^{f(a)}$
\vskip 0.5cm

Assuming $HFS$ extended with {\em Urelements} (i.e. objects not having any
elements), we will use a recursively built {\em rose tree} for Hereditarily
Finite Sets, where {\em Urelements} are represented as Natural Numbers in the
interval {\tt [0,Ulimit-1]} and sets are represented as Prolog lists without
duplicated elements. The change to Ackermann's mapping, to accomodate
a finite number of Urelements in $[0..ulimit-1]$ is a follows:

\vskip 0.5cm
$f_{ulimit}(x)$ = {\tt if} $x<ulimit$ {\tt then} $x$ {\tt else} $ulimit+\sum_{a
\in x}2^{f_{ulimit}(a)}$ 
\vskip 0.5cm

\begin{prop}
For $ulimit \in Nat$ the function $f_{ulimit}$ is a bijection from $Nat$ to
$HFS$ with Urelements in $[0..ulimit-1]$.
\end{prop}

The proof follows from the fact that no sets map to values smaller than
$ulimit$ and that Urelements map into themselves.
Note that if no Urelements are used, we
obtain the ``pure" $HFS$ universe with the empty set represented as {\tt[]} and mapped to {\tt 0}.

First, let's note that Ackermann's encoding can be seen as the recursive
application of a bijection {\tt set2nat} from finite subsets of $Nat$ to $Nat$, 
that associates to a set of (distinct!) natural numbers a (unique!) natural
number, defined as follows:
\begin{code}
set2nat(Xs,N):-set2nat(Xs,0,N).

set2nat([],R,R).
set2nat([X|Xs],R1,Rn):-R2 is R1+(1<<X),set2nat(Xs,R2,Rn).
\end{code}
In fact, {\tt set2nat} maps a set of exponents of 2 (implemented as
bitshifts) to the associated sum of powers of 2. 
With this representation, Ackermann's encoding from $HFS$ to
$Nat$ {\tt hfs2nat} (parameterized by a fixed default\_ulimit) value, becomes:
\begin{code} 
hfs2nat(N,R):-default_ulimit(D),hfs2nat_(D,N,R).  

hfs2nat_(Ulimit,N,R):-integer(N),!,N>=0,N<Ulimit,!,R=N.
hfs2nat_(Ulimit,Ts,R):-
  maplist(hfs2nat_(Ulimit),Ts,T),
  set2nat(T,R0),
  R is R0+Ulimit.

default_ulimit(0).
\end{code}
Note that to ensure that {\tt hfs2nat\_} is a bijection we have shifted
the result {\tt R0} in the second clause by {\tt Ulimit}, shuch that codes
for sets will always be larger or equal to {\tt Ulimit}. 

To obtain the inverse of the Ackerman encoding, let's first define the 
inverse {\tt nat2set} of the bijection {\tt set2nat}. It decomposes a 
natural number into a list of exponents of 2 (seen as bit positions 
equaling 1 in its bitstring representation, in increasing order).
\begin{code}
nat2set(N,Xs):-findall(X,nat2element(N,X),Xs).

nat2element(N,K):-nat2el(N,0,K).

nat2el(N,K1,Kn):-
  N>0, B is /\(N,1), N1 is N>>1,
  nat2more(B,N1,K1,Kn).

nat2more(1,_,K,K).
nat2more(_,N,K1,Kn):-K2 is K1+1,nat2el(N,K2,Kn).
\end{code}
\begin{comment}
Note that we based {\tt nat2set} on the nondeterministic predicate
{\tt nat2element} that enumerates elements of a set represented as
natural numbers, one at a time.

Such predicates tend to be highly reusable
without building intermediate data structures that need to be decomposed 
later, while the list view of the set is always readily available using 
{\tt findall}.
\end{comment}

The inverse of the (bijective) Ackermann encoding, with urelements 
in the interval {\tt [0,Ulimit-1]} is defined as follows:
\begin{code}
nat2hfs_(Ulimit,N,R):-N>=0,N<Ulimit,!,R=N.
nat2hfs_(Ulimit,N,R):-N>=Ulimit,
  N0 is N-Ulimit,
  nat2set(N0,Ns),
  maplist(nat2hfs_(Ulimit),Ns,R).
\end{code}
We can now define
\begin{code}  
nat2hfs(N,R):-default_ulimit(D),nat2hfs_(D,N,R).
\end{code}
where the constant given by {\tt default\_ulimit/1} controls the initial 
segment of $Nat$ to be mapped to {\em Urelements}. We will assume
in {\tt default\_ulimit(0)} unless specified otherwise. Note also
that we shif back {\tt N0} by {\tt Ulimit} to ensure that {\tt nat2hfs\_}
accurately reverses the action of {\tt hfs2nat\_}.
One can try out {\tt nat2hfs} and its inverse {\tt hfs2nat} and
their parametric variants for {\tt Ulimit=3} as follows:
\begin{verbatim}
?- nat2hfs(42,S),hfs2nat(S,N).
S = [[[]], [[], [[]]], [[], [[[]]]]],
N = 42.

?-  nat2hfs_(3,42,S),hfs2nat_(3,S,N).
S = [0, 1, 2, [1]],
N = 42.
\end{verbatim}

\begin{comment}
Figure \ref{f42} shows the directed acyclic graph obtained by merging shared 
nodes in the {\em rose tree} representation of the $HFS$ associated to 
a natural number (with arrows pointing from sets to their elements).
%\HFIGS{f42}{Hereditarily Finite Sets associated to 42}
%{as a pure $HFS$}{with Urelements 0,1,2}{42}{42u3}
\FIG{f42}{Hereditarily Finite Set associated to 42}{42}
\end{comment}

As both {\tt nat2hfs} and {\tt hfs2nat} are obtained through 
recursive compositions of {\tt nat2set} and {\tt set2nat}, respectively, 
one can generalize the encoding mechanism by replacing these building 
blocks with other bijections with similar properties.

\subsection{Combinatorial Generation as Iteration}
Using the inverse of Ackermann's encoding, the infinite stream $HFS$ 
(with urelements in {\tt [0,Ulimit-1]}) can be generated simply by 
iterating over all natural numbers:
\begin{code}
nat(0).
nat(N):-nat(N1),N is N1+1.

iterative_hfs_generator(HFS):-default_ulimit(D),hfs_with_urelements(D,HFS).

hfs_with_urelements(Ulimit,HFS):-nat(N),nat2hfs_(Ulimit,N,HFS).
\end{code}
\begin{verbatim}
?- iterative_hfs_generator(HFS).
HFS = [] ;
HFS = [[]] ;
HFS = [[[]]] ;
HFS = [[], [[]]] ;
HFS = [[[[]]]] ;
HFS = [[], [[[]]]] ;
HFS = [[[]], [[[]]]] ;
...
\end{verbatim}

\subsection{Generating the Stream of Hereditarily Finite Sets Directly} \label{gen}
To fully appreciate the elegance and simplicity of the combinatorial generation 
mechanism described previously, we will also provide a ``hand-crafted" recursive 
generator for $HFS$. The reader will notice that it uses Prolog's
``backtracking over infinite streams of answers" capability in an essential
way. And arguably, that in a language without such features 
the algorithm is likely to get significantly more intricate.

If $P(x)$ denotes the powerset of $x$, the Universe of Hereditarily Finite Sets $HFS$ 
is constructed inductively as follows:
\begin{enumerate}
\item the empty set \{\} is in $HFS$
\item if $x$ is in $HFS$ then the union of its power sets $P^k(x)$ is in $HFS$
\end{enumerate}
\noindent To implement in Prolog a simple $HFS$ generator, conforming this 
definition, we start with a powerset predicate, working with sets represented as
lists:

\begin{code}
all_subsets([],[[]]).
all_subsets([X|Xs],Zss):-all_subsets(Xs,Yss),extend_subsets(Yss,X,Zss).

extend_subsets([],_,[]).
extend_subsets([Ys|Yss],X,[Ys,[X|Ys]|Zss]):-extend_subsets(Yss,X,Zss).
\end{code}  
We can now backtrack over the infinite stream of ``pure" hereditarily finite
sets, one level at a time:
\begin{code}
hfs_generator(NewSet):-nat(N),hfs_level(N,NewSet).
  
hfs_level(N,NewSet):-N1 is N+1,
  subsets_at_stage(N1,[],Hss1),subsets_at_stage(N,[],Hss),
  member(NewSet,Hss1),not(member(NewSet,Hss)).
  
subsets_at_stage(0,X,X).
subsets_at_stage(N,X,Xss):-N>0,N1 is N-1,
  all_subsets(X,Xs),
  subsets_at_stage(N1,Xs,Xss).
\end{code}
Note that redundant generation is avoided by 
keeping only new sets generated at each stage. Note also that
{\tt hfs\_generator} produces the same stream of answers as
{\tt iterative\_hfs\_generator}.
\begin{verbatim}
?- hfs_generator(HFS),hfs2nat(HFS,N).
HFS = [], N = 0 ;
HFS = [[]], N = 1 ;
HFS = [[[]]], N = 2 ;
HFS = [[], [[]]],
...
\end{verbatim}
\subsection{Encoding Hypergraphs}
By limiting recursion to one level in Ackermann's encoding, we can 
derive a bijective encoding of {\em hypergraphs} (also called {\em set systems}), 
represented as sets of sets of {\em Urelements}
\begin{code}
nat2hypergraph(N,Nss):-nat2set(N,Ns),maplist(nat2set,Ns,Nss).

hypergraph2nat(Nss,N):-maplist(set2nat,Nss,Ns),set2nat(Ns,N).
\end{code}
as shown in the following example:
\begin{verbatim}
?- nat2hypergraph(2008,Nss),hypergraph2nat(Nss,N).
Nss = [[0, 1], [2], [1, 2], [0, 1, 2], [3], [0, 3], [1, 3]],
N = 2008.
\end{verbatim}
Like in the case of combinatorial generation of $HFS$, the infinite 
stream of hypergraphs becomes simply
\begin{verbatim}
?-nat(N),nat2hypergraph(N,Nss).
\end{verbatim}
Note also that a hypothetical application using integers, finite sets and hypergraphs 
can now {\em internally share the same data representation} - for instance
arbitrary length integers - opening the doors for a form of generalized memoing
mechanism.

In the following sections we will think about Ackermann's encoding and its 
inverse as {\em Functors} in Category Theory \cite{PIERCE91}, transporting 
various operations from Natural Numbers to Hereditarily Finite Sets and back.

\section{``Shapeshifting'' 
  between $Nat$ and $HFS$ with Fold operators and Functors}
  
\label{functors} Given the {\em rose tree} structure of $HFS$, a natural {\tt fold} 
operation \cite{DBLP:conf/tphol/NipkowP05} can be defined 
on them as a higher order predicate:
\begin{code}
hfold(_,G,N,R):- integer(N),!,call(G,N,R).
hfold(F,G,Xs,R):-maplist(hfold(F,G),Xs,Rs),call(F,Rs,R).
\end{code}
For instance, it can count how many sets occur in a given $HFS$, as follows:
\begin{code}
hsize(HFS,Size):-hfold(hsize_f,hsize_g,HFS,Size).

hsize_f(Xs,S):-sumlist(Xs,S1),S is S1+1.
hsize_g(_,1). 
\end{code}
Note that recursing over {\tt nat2set} has been used to build a member 
of $HFS$ from a member of $Nat$. After lifting {\tt nat2set} to a generic
transformer predicate $T$, we can combine it with the action of a {\tt fold}
operator working directly on natural numbers (with urelements in {\tt
[0,Ulimit-1]}) as shown in the predicate {\tt gfold/6}:
\begin{code}
gfold(_,G,Ulimit,_,N,R):- integer(N),N<Ulimit,!,call(G,N,R).
gfold(F,G,Ulimit,T,N,R):-
  call(T,N,TransformedN),
  maplist(gfold(F,G,Ulimit,T),TransformedN,Rs),
  call(F,Rs,R).
\end{code}
We can now instantiate {\tt gfold} to apply the transformer {\tt nat2set}:
\begin{code}
nfold(F,G,Ulimit,N,R):-gfold(F,G,Ulimit,nat2set,N,R).
nfold1(F,G,N,R):-default_ulimit(D),nfold(F,G,D,N,R).
\end{code}
Note that this can be seen as a form of implicit 
{\em deforestation by prevention}, 
equivalent to deforestation through program
transformation in functional languages \cite{journals/tcs/Wadler90}. 
For instance, {\tt nfold} allows counting the elements contained in the $HFS$ 
representation of a number, without actually building the $HFS$ tree
as in:
\begin{code}
nsize(N,R):-default_ulimit(Ulimit),nsize(Ulimit,N,R).
nsize(Ulimit,N,R):-nfold(hsize_f,hsize_g,Ulimit,N,R).
\end{code}
Note also that {\tt nsize} can be seen as a {\em structural complexity} measure
associated to a Natural Number or bitstring.

The action of the Ackermann encoding as a $Functor$
from $HFS$ to $Nat$ on morphisms (seen as functions on a list of arguments) is
defined as follows:
\begin{code}
toNat(F,Hs,R):-maplist(hfs2nat,Hs,Ns),call(F,Ns,N),nat2hfs(N,R).
\end{code}
The same, acting on 1 and 2 argument operations is:
\begin{code}
toNat1(F,X,R):-hfs2nat(X,N),call(F,N,NR),nat2hfs(NR,R).
  
toNat2(F,X,Y,R):-
  hfs2nat(X,NX),hfs2nat(Y,NY),
    call(F,NX,NY,NR),
  nat2hfs(NR,R).
\end{code}
The inverse Ackermann encoding from $Nat$ to $HFS$ seen as Functor is:
\begin{code}
toHFS(F,Ns,N):-maplist(nat2hfs,Ns,Hs),call(F,Hs,H),hfs2nat(H,N).
\end{code}
with variants acting on a 1 and 2 argument functions:
\begin{code}
toHFS1(F,X,R):-nat2hfs(X,N),call(F,N,NR),hfs2nat(NR,R).
  
toHFS2(F,X,Y,R):-
  nat2hfs(X,NX),nat2hfs(Y,NY),
  call(F,NX,NY,NR),hfs2nat(NR,R).
\end{code}

Using these functors we can define the equivalent of union,
intersection, difference, ordered pair, cartesian product,
powerset, adduction \cite{kaye07,DBLP:journals/mlq/Kirby07}, etc., 
on natural numbers seen as sets. We can also transport 
from $Nat$ to $HFS$, 
operations like successor, sum, product, equality,
with the practical idea in mind that one can pick the most efficient 
(or the simpler to implement) of the two representations.
 
\section{Pairing Functions} \label{pairings}

{\em Pairings} are bijective functions $Nat \times Nat \rightarrow Nat$.

We refer to  \cite{DBLP:journals/tcs/CegielskiR01} for a typical use in the 
foundations of mathematics and to \cite{DBLP:conf/ipps/Rosenberg02a} for an 
extensive study of various pairing functions and their computational properties. 

On top of the ``set operations" defined in subsection \ref{functors} on $Nat$, 
the classic Kuratowski {\em ordered pair}
\begin{math}
(a,b) = \{\{a\},\{a,b\}\}
\end{math}
can be easily implemented.
However, the Kuratowski pair only provides an injective 
function $Nat \times Nat \rightarrow Nat$, resulting in fast 
growing integers very quickly, i.e. for $X,Y \in \{0,1,2,3\}$ the
generated sequence is:
\begin{verbatim}
2,10,34,514,12,4,68,1028,48,80,16,4112,768,1280,4352,256
\end{verbatim}

\subsection{Cantor's Pairing Function}
We can do better by borrowing some interesting 
pairing functions defined on natural numbers. 
Starting from Cantor's pairing function
\begin{code}
cantor_pair(K1,K2,P):-P is (((K1+K2)*(K1+K2+1))//2)+K2.
\end{code}
bijections from $Nat \times Nat$ to $Nat$ have been used for various proofs 
and constructions of mathematical objects 
\cite{robinson50,robinsons68b,DBLP:journals/tcs/CegielskiR01}. 

One can see that this time that the range is more compact, for $X,Y \in
\{0,1,2,3\}$ the sequence is:
\begin{verbatim}
0,2,5,9,1,4,8,13,3,7,12,18,6,11,17,24
\end{verbatim}

\begin{comment}
Note however, that the inverse of Cantor's pairing function involves potentially
expensive floating point operations that are also likely to loose precision
for arbitrary length integers:
\begin{code}
cantor_unpair(Z,K1,K2):-I is floor((sqrt(8*Z+1)-1)/2),
  K1 is ((I*(3+I))//2)-Z,
  K2 is Z-((I*(I+1))//2).
\end{code}
\end{comment}
  
\subsection{An Efficient Pairing Function: BitMerge} \label{BitMerge}

We will introduce here an unusually simple pairing function (that we
have found out recently as being the same as the one in defined 
in Steven Pigeon's PhD thesis on Data Compression \cite{pigeon}, page 114).

The predicate {\tt bitmerge\_pair} implements a bijection from $Nat \times
Nat$ to $Nat$ that works by splitting a number's big endian bitstring
representation into odd and even bits, while its inverse {\tt to\_pair} blends
the odd and even bits back together. We will provide here an ``elementary''
implementation using exclusively bitshifting and addition operations, while
abstracting away the set view of a natural number through the generator {\tt
nat2element} defined in subsection \ref{ack}.

\begin{code}
bitmerge_pair(A,B,P):-up0(A,X),up1(B,Y),P is X+Y.

bitmerge_unpair(P,A,B):-down0(P,A),down1(P,B).

even_up(A,R):-nat2element(A,X),E is X<<1,R is 1<<E.
odd_up(A,R):-nat2element(A,X),E is 1+(X<<1),R is 1<<E.
even_down(A,R):-nat2element(A,X),even(X),E is X>>1,R is 1<<E.
odd_down(A,R):-nat2element(A,X),odd(X),E is (X>>1), R is 1<<E.

even(X):- 0 =:= /\(1,X).
odd(X):- 1 =:= /\(1,X).

up0(A,P):-findall(R,even_up(A,R),Rs),sumlist(Rs,P).
up1(A,P):-findall(R,odd_up(A,R),Rs),sumlist(Rs,P).
down0(A,X):-findall(R,even_down(A,R),Rs),sumlist(Rs,X).
down1(A,X):-findall(R,odd_down(A,R),Rs),sumlist(Rs,X).
\end{code}

Note that {\tt up} operations insert 0's in each even/odd position of the
bitstrings while {\tt down} operations remove even/odd bits while keeping
odd/even bits, as shown in the following example with bitstrings aligned:
\begin{verbatim}
?- bitmerge_unpair(2008,X,Y),bitmerge_pair(X,Y,Z).
X = 60,
Y = 26,
Z = 2008.
% 2008:[0, 0, 0, 1, 1, 0, 1, 1, 1, 1, 1]
%   60:[      0,    1,    1,    1,    1]
%   26:[   0,    1,    0,    1,    1   ]
\end{verbatim}

Note also the significantly more compact packing, compared to Kuratowski 
pairs, and, like Cantor's pairing function, similar growth in both arguments.
\begin{comment}
\begin{verbatim}
?- between(0,15,N),bitmerge_unpair(N,A,B),
   write(N:(A,B)),put(32),fail;nl.
0: (0, 0) 1: (1, 0) 2: (0, 1) 3: (1, 1) 
4: (2, 0) 5: (3, 0) 6: (2, 1) 7: (3, 1)
8: (0, 2) 9: (1, 2) 10: (0, 3) 11: (1, 3) 
12: (2, 2) 13: (3, 2) 14: (2, 3) 15: (3, 3)

?- between(0,3,A),between(0,3,B),bitmerge_pair(A,B,N),
   write(N:(A,B)),put(32),fail;nl.
0: (0, 0) 2: (0, 1) 8: (0, 2) 10: (0, 3) 
1: (1, 0) 3: (1, 1) 9: (1, 2) 11: (1, 3) 
4: (2, 0) 6: (2, 1) 12: (2, 2) 14: (2, 3) 
5: (3, 0) 7: (3, 1) 13: (3, 2) 15: (3, 3)
\end{verbatim}
\end{comment}

It is convenient to see pairing/unpairing as one-to-one
functions from/to the underlying language's ordered pairs:
\begin{code}
bitmerge_pair(X-Y,Z):-bitmerge_pair(X,Y,Z).

bitmerge_unpair(Z,X-Y):-bitmerge_unpair(Z,X,Y).
\end{code}
This view as one-argument functions will allow using them
in operations like {\tt maplist}.

\section{Powersets, Ordinals and Choice Functions} \label{ord}
A concept of (finite) {\em powerset} can be associated to a 
number $n \in Nat$  by computing the powerset of the $HFS$ 
associated to it, using the {\tt toHFS1} functor:

\begin{code}
nat_powset(N,PN):-toHFS1(all_subsets,N,PN).
\end{code}
\begin{comment}
or, directly, as in \cite{abian78}:
\begin{code}
%nat_powset_alt i = product (map (\k->1+(exp2 . exp2) k) (nat2set i)) 
\end{code}
\end{comment}

The von Neumann {\em ordinal} associated to a $HFS$, defined with 
interval notation as $\lambda=[0,\lambda)$, is implemented by the 
function {\tt hfs\_ordinal}, simply by transporting it from $Nat$:
\begin{code}
hfs_ordinal(0,[]).
hfs_ordinal(N,Os):-N>0,N1 is N-1,findall(I,between(0,N1,I),Is),
  maplist(hfs_ordinal,Is,Os).
 
nat_ordinal(N,OrdN):-hfs_ordinal(N,H),hfs2nat(H,OrdN).
\end{code}

The following example shows the {\em transitive structure} of a 
von Neumann ordinal's set representation. %$$ (see Fig. \ref{f4}). 
It also shows its fast growing $Nat$ encoding ($4 \rightarrow 2059$) 
which can be seen as a somewhat unusual injective embedding of 
finite ordinals in $Nat$, seen as the set of finite cardinals.

\begin{verbatim}
?- hfs_ordinal(4,H),nat_ordinal(4,N),write(N:H),nl.
2059:[[], [[]], [[], [[]]], [[], [[]], [[], [[]]]]]
\end{verbatim}

\begin{comment}
\HFIGS{f4}{4 and its associated ordinal}{as a pure $HFS$}
{its associated ordinal 2059}{4}{2059}
\end{comment}

Finally, a choice function, showing that $Nat$ with the structure 
borrowed from $HFS$ is actually a model for the Axiom of Choice, 
is implemented as an encoding of pairs of sets and their first 
elements with our compact $Nat \times Nat \rightarrow Nat$ 
pairing function {\tt bitmerge\_pair}:
\begin{code}
nat_choice_fun(N,CFN):-nat2set(N,Es),
  maplist(nat2set,Es,Ess),maplist(choice_of_one,Ess,Hs),
  maplist(bitmerge_pair,Es,Hs,Ps),set2nat(Ps,CFN).

choice_of_one([X|_],X).
\end{code}
As {\em even} numbers represent sets that do not contain the 
empty set as an element, we  compute $Nat$ representations 
of the choice function as follows:
\begin{verbatim}
?- maplist(nat_choice_fun,[0,2,4,6,8,10,12,14,16],Funs).
Funs = [0, 2, 64, 66, 32, 34, 96, 98, 16777216].
\end{verbatim}

\section{Directed Graph Encodings} \label{graphs}
Directed Graphs are equivalent to binary relations seen as 
sets of ordered pairs. Equivalently, they can also be seen as 
vertices paired with lists of vertices of adjacent outgoing edges.

\subsection{Directed Acyclic Graph representations for $HFS$} \label{graphrep}
The tree representation of $HFS$ can be seen as a {\em set} 
of edges, oriented to describe either set membership $\in$ 
or its transpose, set containment $\ni$:
\begin{code}
nat2memb(N,XY):-default_ulimit(D),nat2memb(D,N,XY).
nat2memb(Ulimit,N,X-Y):-nat2contains(Ulimit,N,Y-X).

nat2contains(N,XY):-default_ulimit(D),nat2contains(D,N,XY).
nat2contains(Ulimit,N,E):-
  N>=Ulimit,
  N0 is N-Ulimit,
  nat2element(N0,X),
  ( E=N-X
  ; nat2contains(Ulimit,X,E)
  ).
\end{code}
The following examples show how the two predicates work.
%in accordance to Fig. \ref{f42}, section \label{ack}:
\begin{verbatim}
?- findall(X,nat2memb(42,X),Xs),sort(Xs,S),write(S),nl.
[0-1, 0-3, 0-5, 1-2, 1-3, 1-42, 2-5, 3-42, 5-42]

?- findall(X,nat2contains(42,X),Xs),sort(Xs,S),write(S),nl.
[1-0, 2-1, 3-0, 3-1, 5-0, 5-2, 42-1, 42-3, 42-5]
\end{verbatim}
The pair representation of $\in$ and its inverse $\ni$ can be turned
directly into an actual graph data type (using SWI-Prolog's graph and
associative list libraries). Given that our sets are well-founded, this graph
is a directed acyclic graph.
\begin{code}
nat2cdag(Ulimit,N,G):-
  findall(E,nat2contains(Ulimit,N,E),Es),
  vertices_edges_to_ugraph([],Es,G).

nat2mdag(Ulimit,N,G):-
  findall(E,nat2memb(Ulimit,N,E),Es),
  vertices_edges_to_ugraph([],Es,G).  
\end{code}
\begin{verbatim}
?- nat2cdag(1,42,G).
G=[0-[], 1-[], 2-[0], 3-[1], 5-[2], 42-[0, 3, 5]].

?- nat2mdag(1,42,G).
[0-[2, 42], 1-[3], 2-[5], 3-[42], 5-[42], 42-[]].
\end{verbatim}
However, as the associated $HFS$ is usually sparse in case of large
integers, we can think about compressing it to a canonical form.
We can achieve this by replacing its $n$ distinct vertex numbers
with smaller integers in $[0,n-1]$, by progressively building a map 
describing this association. Given the fast growth of nodes
from a level to another in a $HFS$ tree,
it makes sense to map the largest nodes to small integers first,
as shown in the predicate {\tt to\_dag}:
\begin{code} 
to_dag(N,NewG):-
  findall(E,nat2contains(0,N,E),Es),
  vertices_edges_to_ugraph([],Es,G),
  vertices(G,Rs),reverse(Rs,Vs),
  empty_assoc(D),remap(Vs,0-D,_RVs,KD),remap(Es,KD,REs,_NewKD),
  vertices_edges_to_ugraph([],REs,NewG).

remap(Xs,Rs):-empty_assoc(D),remap(Xs,0-D,Rs,_KD).

remap([],KD,[],KD).
remap([X|Xs],KD1,[A|Rs],KD3):-integer(X),!,
  assoc(X,A,KD1,KD2),
  remap(Xs,KD2,Rs,KD3).
remap([X-Y|Xs],KD1,[A-B|Rs],KD4):-
  assoc(X,A,KD1,KD2),assoc(Y,B,KD2,KD3),
  remap(Xs,KD3,Rs,KD4).
  
assoc(X,R,K-D,KD):-get_assoc(X,D,A),!,R=A,KD=K-D.
assoc(X,K,K-D,NewK-NewD):-NewK is K+1,put_assoc(X,D,K,NewD).
\end{code}
Note that this construction assumes sets with no urelements, and the root of the
graph (represented as $0$) is the original natural number $n$ 
from which the $HFS$ has been built.
\begin{verbatim}
?- to_dag(42,G).
G = [0-[1, 2, 4], 1-[3, 5], 2-[4, 5], 3-[4], 4-[5], 5-[]]
\end{verbatim}

An interesting question arises at this point. 
{\em Can we rebuild a natural number from its directed acyclic
graph representation, assuming no labels are available, except 0?} 

The answer is yes, and the key idea here is to apply {\tt set2nat}
recursively in a way similar to the implementation of {\tt hfs2nat} 
and rebuild
the values in each vertex while progressing 
towards the root to recover the original value of $n \in Nat$,
as shown in {\tt from\_dag}:
\begin{code}   
from_dag(G,N):-vertices(G,[Root|_]),compute_decoration(G,Root,N).

compute_decoration(G,V,Ds):-neighbors(V,G,Es),compute_decorations(G,Es,Ds).
  
compute_decorations(_,[],0).
compute_decorations(G,[E|Es],N):-
  maplist(compute_decoration(G),[E|Es],Ds),
  set2nat(Ds,N).
\end{code}
\begin{verbatim}
?- to_dag(42,G),from_dag(G,N).
G = [0-[1, 2, 4], 1-[3, 5], 2-[4, 5], 3-[4], 4-[5], 5-[]], N = 42
\end{verbatim}

After implementing this predicate, we have found that it closely 
follows the {\em decoration} functions used in Aczel's book \cite{aczel88}, 
and renamed it {\tt compute\_decoration}. In the simpler case of the $HFS$ universe, 
with our well-founded sets represented as DAGs, the existence and unicity of the 
result computed by {\tt from\_dag} follows immediately from the 
Mostowski Collapsing Lemma \cite{aczel88}.

\subsection{Encodings of Directed Graphs as Natural Numbers} \label{digraphs}
Hypersets \cite{aczel88} are defined by replacing the 
Foundation Axiom with the AntiFoundation axiom. 
Intuitively this means that the $\in$-graphs can 
be cyclical \cite{barwise96}, provided that 
they are minimized through 
{\em bisimulation equivalence} \cite{DBLP:conf/cav/DovierPP01}.
We have not (yet) found an elegant encoding of hereditarily finite 
hypersets as natural numbers, similar to Ackerman's encoding. 
The main difficulty seems related to the fact that hypersets are 
modeled in $HFS$ as equivalence classes with respect to 
bisimulation \cite{aczel88,barwise96,DBLP:journals/tplp/PiazzaP04}. 
Toward this end, an easy first step seems to find a bijection from 
directed graphs (with no isolated vertices, corresponding to their
 view as binary relations), to $Nat$:
\begin{code}
nat2digraph(N,G):-nat2set(N,Ns),
  maplist(bitmerge_unpair,Ns,Ps),
  vertices_edges_to_ugraph([],Ps,G).
  
digraph2nat(G,N):-edges(G,Ps),
  maplist(bitmerge_pair,Ps,Ns),
  set2nat(Ns,N).
\end{code}
With digraphs represented as lists of edges, this bijection 
works as follows:
\begin{verbatim}
2 ?- nat2digraph(255,G),digraph2nat(G,N).
G = [0-[0, 1], 1-[0, 1], 2-[0, 1], 3-[0, 1]]
N = 255.

3 ?- nat2digraph(2008,G),digraph2nat(G,N).
G = [0-[2, 3], 1-[1, 2], 2-[0, 1], 3-[1]],
N = 2008.
\end{verbatim}
\begin{comment}
The resulting graphs are pictured in Figure \ref{n2g}.
\HFIGS{n2g}{Digraph Encodings}{2008 as a digraph}{255 as a
digraph}{g2008}{g255} 
\end{comment}
As usual
\begin{verbatim}
?-nat(N),nat2digraph(N,G).
\end{verbatim}
provides a combinatorial generator for the infinite stream of 
directed acyclic graphs.

\subsection{Duality}
Idempotent operations like reversing the sense of the edges in a
digraph (transpose/2 in SWI-Prolog) induce $Nat \rightarrow Nat$ bijections:
\begin{code}
transpose_nat(N,TN):-nat2digraph(N,G),transpose(G,T),digraph2nat(T,TN).
\end{code}
\begin{verbatim}
1?- maplist(transpose_nat,[0,1,2,3,4,5,6,7],Ts),
    maplist(transpose_nat,Ts,Ns).
Ts = [0, 1, 4, 5, 2, 3, 6, 7],
Ns = [0, 1, 2, 3, 4, 5, 6, 7].
\end{verbatim}
A more interesting form of duality arises when we interchange the
$\in$ and $\ni$ relations themselves in $HFS$. 
Intuitively, it corresponds
to the fact that intensions/concepts 
would become the building blocks of the theory, provided 
that something similar to the {\em axiom of extensionality} holds. 
In comments related to Russell's type theory \cite{russelcrit} 
pp. 457-458 G\"{o}del mentions an {\em axiom of intensionality} with 
the intuitive meaning that ``different definitions belong to
different notions". G\"{o}del also notices the duality between 
``no two different properties belong to exactly the same things" 
and ``no two different things have exactly the same properties"  
but warns that contradictions in a simple type theory would result 
if such an axiom is used non-constructively. We will leave
as a topic for future research to investigate various 
aspects of $\in$ / $\ni$ duality in $HFS$, in correlation 
with Natural Number their encodings.

\section{Related work} \label{related}
Natural Number encodings of Hereditarily Finite Sets have triggered 
the interest of researchers in fields ranging from Axiomatic Set Theory 
and Foundations of Logic to Complexity Theory and Combinatorics
\cite{finitemath,kaye07,DBLP:journals/mlq/Kirby07,DBLP:journals/jsyml/Booth90,DBLP:conf/foiks/LeontjevS00}. 
Graph representations of sets and hypersets based on the variants of the 
Anti Foundation Axiom have been studied extensively in \cite{aczel88,barwise96}.
Computational and Data Representation aspects of Finite Set Theory and 
hypersets have been described in a logic programming context in 
\cite{dovier00comparing,DBLP:journals/tplp/PiazzaP04,DBLP:conf/cav/DovierPP01}. 
Pairing functions have been used work on decision problems as early 
as \cite{pepis,robinson50,robinsons68b}. 

\section{Conclusion and Future Work} \label{concl}
Implementing with relative ease the encoding techniques
typically used only in the foundations of mathematics recommends logic programming
languages as effective tools for experimental mathematics.

While focusing on the ability to ``shapeshift" between different data types, 
with the intent of internally 
sharing possibly heterogeneous data representations, we have described a variety 
of isomorphisms between mathematically interesting data structures, all centered 
around encodings as Natural Numbers. The possibility of sharing significant
common parts of HFS-represented integers could be used in implementing shared 
stores for arbitrary length integers. Along the same lines, another application 
would be data compression using 
some ``information theoretically minimal" variants of the graphs in 
subsection \ref{graphrep}, from which larger, $HFS$ and/or 
natural numbers can be rebuilt.
\bibliographystyle{plain}
%\bibliography{INCLUDES/theory,tarau,INCLUDES/proglang,INCLUDES/biblio,INCLUDES/syn}

\begin{comment}
\appendix
\section{APPENDIX}
To make the code in the paper fully self contained, 
we list here some auxiliary functions.

\paragraph{Output Predicates}
The following predicate prints out a $HFS$ with {\em Urelements}.
\begin{code}
setShow(S):-gshow(S,"{,}"),nl.

gshow(0,[L,_C,R]):-put(L),put(R).
gshow(N,_):-integer(N),N>0,!,write(N).
gshow(Hs,[L,C,R]):-put(L),gshow_all(Hs,[L,C,R]),put(R).

gshow_all([],_).
gshow_all([H],LCR):-gshow(H,LCR).
gshow_all([H,G|Hs],[L,C,R]):-
  gshow(H,[L,C,R]),
  ([C]\=="~"->put(C);true),
  gshow_all([G|Hs],[L,C,R]).
  
test:-
  G=[0-[1, 2, 5, 6, 7], 1-[7, 9], 2-[7, 10], 3-[7], 
     4-[8, 10],5-[8, 9], 6- [8], 7-[9], 8-[9], 9-[10], 10-[]],
  from_dag(G,N),
  to_dag(N,G1),
  from_dag(G1,N2),
  write(N+G),nl,nl,
  write(N2+G1),nl,nl.

c:-['pSET.pro'].
  
\end{code}
\end{comment}
\end{document}